# Interaction of neutrons with alpha particles:
# A tribute to Heinz Barschall[1]

Bernard Hoop

Science Faculty, Riley College of Education, Walden University, Minneapolis, MN 55401, USA
bernard.hoop@waldenu.edu

**Abstract**

As a tribute to our teacher and mentor on the occasion of his centennial celebration, we provide a brief historical overview and a summary of sustained interest in the topic of interaction of neutrons with alpha particles

## 1 Introduction

Henry Herman Barschall (1915-1997), a physics professor at the University of Wisconsin-Madison from 1946 to 1986, known to his students and friends as "Heinz," is internationally known and respected for his research with fast neutrons [1]. One of his earliest contributions is a study of neutrons scattered by alpha particles. Interaction of neutrons with $^4$He occurs in primordial and stellar nucleosynthesis, in nuclear fusion, and is of interest in *ab initio* theory of light ion reactions [2]. Spin-orbit coupling and resonant scattering of neutrons by alpha particle are predominant features.

The angular distribution of fast neutrons scattered by helium was first measured in 1940 by Barschall and Kanner [3] A subsequent paper by Wheeler and Barschall [4] summarized: *"From the observations of Barschall and Kanner on the scattering of 2.5-Mev neutrons in helium is deduced the existence at this energy of a coupling between the spin and orbital motion of the neutron. Less clear-cut evidence points to appreciable interaction between a neutron and alpha-particle of two units of mutual angular momentum, the classical distance of closest approach of which would be $7\times10^{-13}$ cm."* Barschall reminisces [5]: *"The most interesting result was that the scattering of fast neutrons by helium is strongly anisotropic. Wheeler was able to show that the observations could be explained only by assuming that the nuclear spin-orbit coupling was very large, of the order of MeVs…The large spin-orbit coupling was later rediscovered by Maria Mayer and is the basis of the nuclear shell model."* Wheeler reminisces [6]: *"Barschall points out that our work really amounted to the first evidence for the spin-orbit coupling which Maria Mayer and Jensen were to invoke for explaining the order of filling of energy levels in the nucleus."*

In experiments with neutrons, the large nuclear spin-orbit coupling in neutron-alpha scattering has made helium a widely used analyzer of polarized neutrons up to neutron energy of about 20 MeV. The analyzing power is calculated from n-alpha phase shifts. Above neutron energy of 22.064 MeV, (threshold for $^4$He(n,d)$^3$H reaction) and the lowest $J^\pi = 3/2^+$ excited state in $^5$He, the neutron-alpha $D_{3/2}$ phase shift varies rapidly and becomes complex. In addition to a conventional resonance pole that is primarily an elastic resonance in the neutron-$^4$He channel, as well as an elastic resonance in the deuteron-$^3$H channel, a shadow pole is located on a different energy plane that contributes to the large cross section for the $^4$He(n,d)$^3$H reaction [7]. It is the inverse of this reaction that is important in primordial and stellar nucleosynthesis and in applications of nuclear fusion energy.

---

[1] Submitted "by title only" to: Barschall/Haeberli Symposium, Dept. of Physics, University of Wisconsin-Madison, WI, April 11, 2015

# 2 Review

Attempts to characterize this neutron-alpha resonant feature are shown in Figure 1. In each row of Figure 1, from left to right, real part of phase shift (degrees), inelastic parameter, and n+$^4$He scattering amplitude are shown over neutron lab energy 22.0 to 23.6 MeV lab (neutron-alpha c.m. energy 17.5 to 18 MeV.

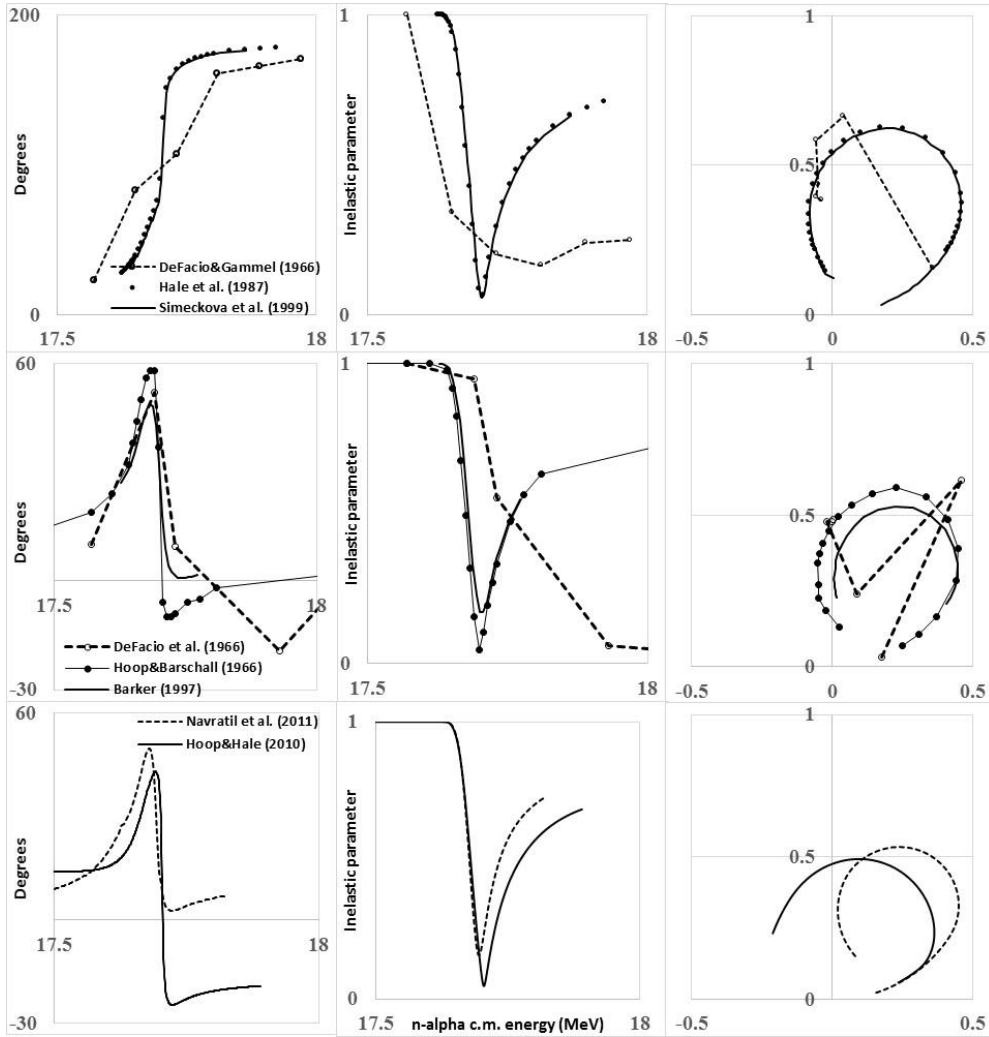

FIGURE 1. Real part of phase shift (left), inelastic parameter (center), and n+$^4$He scattering amplitude (right) are shown over neutron energy 22.0 to 23.6 MeV laboratory (neutron-alpha c.m. energy 17.5 to 18 MeV).

The top row compares results of three different analyses [7-9], each of which yield a $D_{3/2}$ phase shift (real part) that rises through $\pi/2$. Two analyses are in striking agreement. In all cases, corresponding scattering amplitude encloses the center of the unitary circle.

The second row compares results of three other analyses [10-12], for which the $D_{3/2}$ phase shift does not increase through $\pi/2$, and for which corresponding scattering amplitude does not enclose the center.

In another context, G.C. Phillips [13] remarked that *"…the 3/2+ phase shift goes positive, then has a large negative slope…"* and wondered whether there would be *"…any reason to worry about a possible violation of Wigner's theorem ([14]) that a phase shift's slope cannot have too large a negative value."* The third row compares results of two different recent analyses [15, 16] that exhibit this same feature. The analysis in [15] employs a time delay calculation from a comparison of a Breit-Wigner approximation to the resonance data, and suggests a relatively small deuteron partial width, compared to neutron width, which yields a narrow resonant elastic cross section. This analysis suggests that the reported shadow pole width (ca. 0.008 MeV) describes deuteron partial width, whereas the width (ca 0.030 MeV) of peak $^4$He(n,d)$^3$H reaction cross section relative to its maximum describes neutron partial width. Accounting for 0.040-MeV experimental energy spread and cross section scale factor normalization, partial widths from maximum elastic time delay analysis taken as deuteron and neutron partial widths with resonance energy 22.124 MeV, yield the phase shift and scattering amplitude plotted as heavy solid curves in row 3 of Figure 1.

The analysis in [16] is an *ab initio* analysis that does not involve fitting experimental data. These authors point out that their analysis employs similarity-renormalization-group-evolved NN potentials dependent on a cutoff parameter. Therefore, characterization of resonant phase shift and energy may be different, by including three-nucleon (NNN) interaction or different NN potential with modified cutoff parameter.

As reviewed elsewhere [17], other neutron-alpha differential cross sections in the energy range 2 to 30 MeV have been measured by Austin et al. [18], Shamu and Jenkin [19] and Hoop and Barschall [11]. The latter two publications include detailed measurements through the lowest $3/2^+$ state in $^5$He at 22.13 MeV neutron energy, just above the $^4$He(n,d)$^3$H threshold at 22.064 MeV. These measurements employ a recoil particle technique. Namely, Wheeler and Barschall [4] showed that detection and energy measurement of the recoiling associated alpha particle is proportional to the angular distribution of the scattered neutrons in the zero-momentum system if the angular distribution is expressed in terms of the cosine of the center-of-mass scattering angle. This proportionality also holds relativistically.

## 3 Evaluation

Neutron-$^4$He experiments reported by Shamu and Jenkin (SJ) [19] provide total cross section and differential cross section measurements over the neutron energy range 20 to 29 MeV, including closely spaced measurements over the resonance near 22 MeV. In Table III and Figures 6 and 7 of their publication, SJ summarize numerical results of Legendre polynomial fits to their measurements, as well as measurements of angular distributions and excitation functions.

An apparent discrepancy between data in Figure 7 and Table III of SJ was examined, as follows: Neutron-$^4$He angular distributions were digitized from the eight measured excitation functions plotted in Figure 7 of SJ, each consisting of 15 data points over the neutron energy range 21.8 to 22.5 MeV. Each of the 15 angular distributions therefore consists of eight points over cosine c.m. scattering angles +0.35 to -0.88. For comparison, the angular distribution at 22.15 MeV plotted in Figure 6 of SJ was also digitized. From the excitation function plots, three adjacent angular distributions were averaged over the lowest twelve energy points, yielding four angular distributions at neutron energies of 21.995, 22.073, 22.152, and 22.230 MeV. These data are plotted in Figure 2. Angular distribution at 22.321 MeV is included, which is mean of three angular distributions of Hoop and Barschall (HB) [11] from 22.20 to 22.45 MeV. For comparison with the distribution at 22.152 MeV, the angular distribution at 22.15 MeV from SJ Figure 6 is also included in Figure 2.

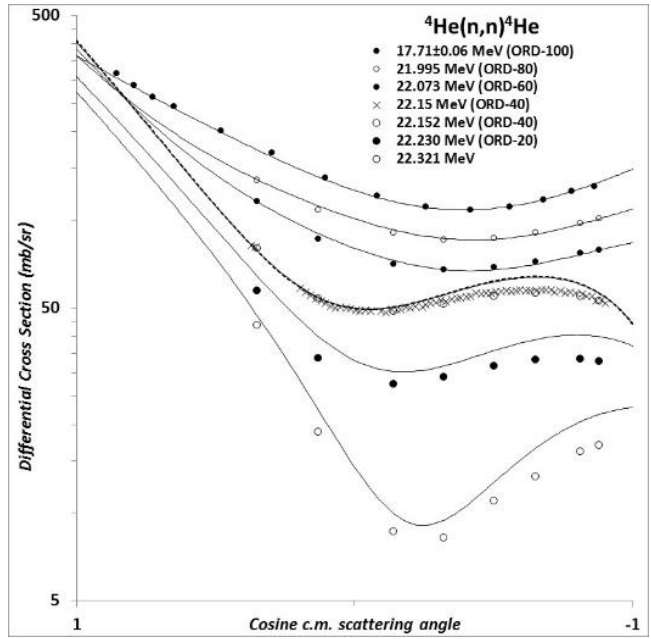

FIGURE 2. Differential cross sections for n-$^4$He elastic scattering.

Angular distributions at each energy are normalized to elastic cross sections taken from a spline-smoothed fit by eye to all measured total cross sections (SJ [19] Figure 2 and Haesner et al. [20] Figure 4), from which was subtracted $^4$He(n,d)$^3$H reaction cross section calculated from tabulated cross section for the inverse reaction [24]. Figure 2 shows solid curves calculated from HB phase shifts [11]. Differential cross section at 17.71 MeV is that of Drosg et al. [17]. Integrated 17.71-MeV differential cross section (881 mb) agrees with Drosg et al. (878±15 mb).

Legendre coefficient A0 vs neutron energy derived from HB and SJ Fig. 7 differ from values in SJ Table III. That is, maximum A0 from HB and SJ occurs at neutron energy $E_n$ < 22.15 MeV, whereas for SJ Table III, maximum occurs at 22.15 MeV as shown in Fig. 3.

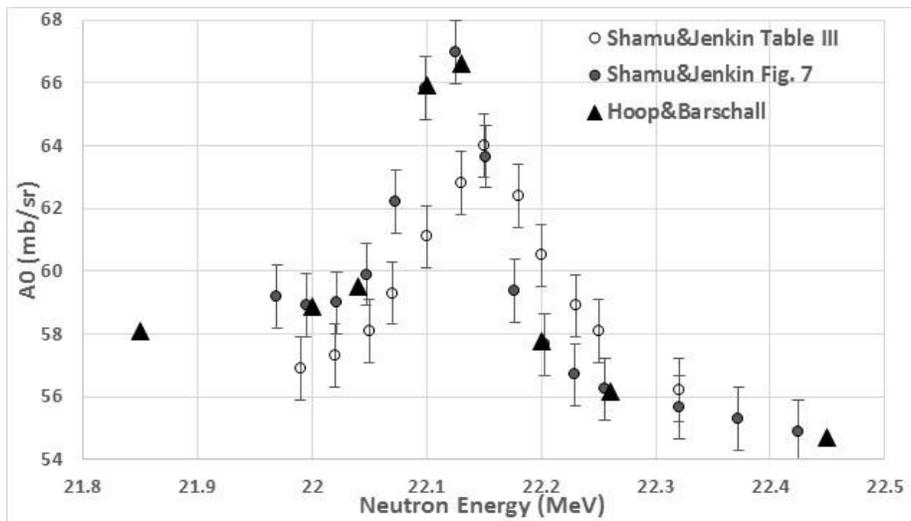

FIGURE 3. Legendre coefficient A0 vs neutron energy, as reported in, and as derived from refs [11] and [19]

With reference to Fig. 2, differential cross sections calculated from HB phase shifts also differ significantly from measured values at 22.15 MeV and above. Difference at 22.15 (and 22.152) MeV is seen in the shape of the angular distribution at back angles, whereas differences at 22.230 and 22.321 MeV appear to be due to an error in normalization of the entire distribution.

In addition to the above differences, resonant phase shift (real part and inelastic parameter) derived directly from all available measured total cross section (SJ and Haesner et al. [20]) and $^4$He(n,d)$^3$H reaction cross section, and corrected for non-resonant background, yield a scattering amplitude that does not encircle the unitary circle origin, in distinction to that found in the analyses of Simeckova et al. [9] and Hale et al [7].

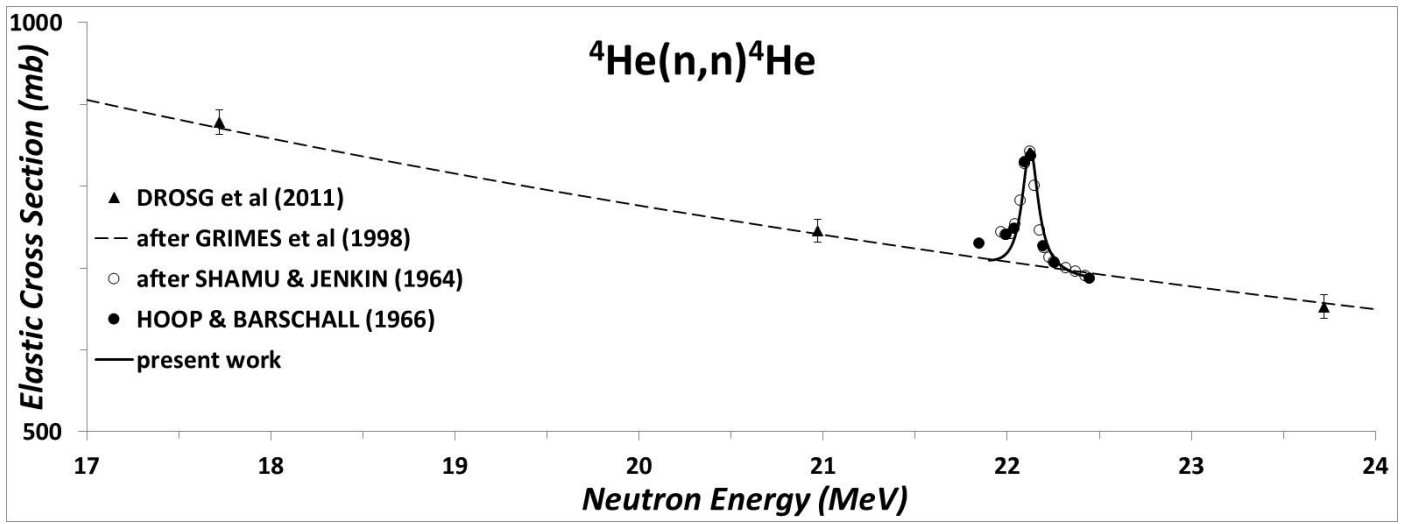

FIGURE 4. Elastic cross section vs neutron energy for n-$^4$He elastic scattering. Solid triangles are from Drosg et al [17]; Solid and open circles are from, respectively, Hoop and Barschall [11] and Shamu and Jenkin [19]; Dashed curve is after Grimes et al. [21] and references therein; Solid curve is a calculation described in the present work.

Three re-evaluated n-$^4$He elastic cross section measurements in the 17- to 24-MeV neutron energy range reported by Drosg et al [17] are shown in Fig. 4. Non-resonant cross section (dashed curve in Fig. 4) is derived from a nuclear optical model after Grimes et al. [21] and references therein, as reported earlier [15]. A calculation of the resonant feature near 22 MeV, discussed in Section 4, below, is in satisfactory agreement with re-evaluated measurements of Shamu and Jenkin [19] and of Hoop and Barschall [11].

# 4 Discussion

**Motivation**

A recent analysis of the $^3$H(d,n)$^4$He reaction by Brown and Hale [22] and by Hale, Brown and Paris [23] employs a two-channel effective field theory, which is further examined using a two-channel, single-level *R*-matrix parameterization. These authors report that the resulting *S* matrix *"…is shown to be identical in these two representations in the limit that R-matrix channel radii are taken to zero"* and note a conversation with Prof. Wigner in 1975 *"…that he was thinking about what R-matrix theory looks like at zero radius."* This observation motivates the present model of a neutron-$^4$He channel radius that goes to zero at an energy corresponding to the energy of unstable particle, $^5$He$^*$, for which time delay in n-$^4$He scattering may provide further evidence.

# $^3$H(d,n)$^4$He reaction from effective field theory

Previous studies of time delay in neutron-$^4$He scattering at d-$^3$H threshold [15] employed an empirical expression for $^4$He(n,d)$^3$H reaction cross section that served as satisfactory parametric representation up to ca. 22.3 MeV neutron laboratory energy, but with no physical justification for the representation. In the present study, cross section for the $^4$He(n,d)$^3$H reaction is calculated via detailed balancing from effective field theory analysis of the d + $^3$H --> n + $^4$He reaction cross section reported by Brown and Hale [22]. The present representation compares favorably with several experimental measurements, as well as with $^4$He(n,d)$^3$H cross section derived via detailed balancing of $^3$H(d,n)$^4$He reported by Bosch and Hale [24], as well as with recent tabulation of Drosg and Otuka [25].

The expression for detailed balancing that relates $^4$He(n,d)$^3$H and $^3$H(d,n)$^4$He cross section is $\sigma_{nd} = 3 \cdot \sigma_{dn}(k_d^2/k_n^2)$, where $\sigma_{nd}$ and $\sigma_{dn}$ are, respectively, $^4$He(n,d)$^3$H and $^3$H(d,n)$^4$He cross sections, and $k_d$ and $k_n$ are relativistic wave numbers in d-$^3$H and n-$^4$He center-of-mass systems[2]

$$k_d^2 = \frac{1}{(\hbar c)^2} \frac{(m_t/c^2)^2 E_d(E_d + 2m_d/c^2)}{(m_d/c^2 + m_t/c^2)^2 + 2m_t/c^2 E_d} \quad \text{and} \quad k_n^2 = \frac{1}{(\hbar c)^2} \frac{(m_\alpha/c^2)^2 E_n(E_n + 2m_n/c^2)}{(m_n/c^2 + m_\alpha/c^2)^2 + 2m_\alpha/c^2 E_n} \quad (1)$$

In their study of effective field theory as a limit of $R$-matrix theory, Hale, Brown and Paris [23] report the following expression for the single-level, two-channel $J^\pi = 3/2^+$ $^3$H(d,n)$^4$He reaction cross section:

$$\sigma_{n,d}^{3/2^+} = \frac{32\pi}{9\hbar v_d} \frac{g_d^2}{4\pi} \frac{g_n^2}{4\pi} \frac{\mu_n}{\hbar^2 c^2} k_n^5 C_0^2(\eta_d) \left| E - E_\lambda - \Delta_d(E) - i\left[\frac{g_d^2}{2\pi} \frac{\mu_d}{\hbar^2 c^2} k_d C_0^2(\eta_d) + \frac{g_n^2}{6\pi} \frac{\mu_n}{\hbar^2 c^2} k_n^5\right] \right|^{-2} \quad (2)$$

Using Eq. (2), we calculate $^3$H(d,n)$^4$He reaction cross section over d-$^3$H c.m. energy range $E = 0$ to 0.3 MeV.

In Eq. (2), we use absolute values of reduced widths ($\gamma_d^2 = 0.324, \gamma_n^2 = 0.0122$ MeV, respectively) and interaction radii ($a_d = 35.48, a_n = 1.767923$ fm, respectively) in the expressions for coupling constants,

$$g_n^2 = -\frac{6\pi\gamma_n^2(\hbar c)^2 a_n^5}{\mu_n} \quad \text{and} \quad g_d^2 = -\frac{2\pi\gamma_d^2(\hbar c)^2 a_d}{\mu_d} \quad (3)$$

reported in [23]. This yields absolute values $g_d^2 = 4\pi \cdot 0.199, g_n^2 = 4\pi \cdot 0.0164$ fm$^3$ MeV$^2$ and fm$^7$ MeV$^2$, respectively, which agree with values reported by Brown and Hale [22] (N.B.: typographical error in units of coupling constants in that publication). We also use $E_* = -0.154$ MeV for the unrenormalized energy of the unstable $^5$He$^*$ particle reported in [22].

---

[2] We use a system of units, where $\hbar = 6.58212 \cdot 10^{-7}$ MeV · fsec , $c = 2.997925 \cdot 10^8$ fm/fsec , $m_n/c^2 = 939.5653$ MeV, $m_\alpha/c^2 = 3727.3791$ MeV, $m_t/c^2 = 2808.9209$ MeV, $m_d/c^2 = 1875.6128$ MeV. Corresponding reduced masses $\mu_d$ and $\mu_n$ for the d-$^3$H and n-$^4$He systems are, respectively, $\mu_d/c^2 = (m_t/c^2 \cdot m_d/c^2)/(m_t/c^2 + m_d/c^2) = 1124.647$ MeV and $\mu_n/c^2 = (m_n/c^2 \cdot m_\alpha/c^2)/(m_n/c^2 + m_\alpha/c^2) = 750.409$ MeV. Reaction Q-value and corresponding neutron laboratory threshold energy $E_{Th}$ for the $^4$He(n,d)$^3$H reaction are $Q = m_d/c^2 + m_t/c^2 - m_n/c^2 - m_\alpha/c^2 = 17.589$ MeV and $E_{Th} = 0.5 \cdot Q \cdot (m_n/c^2 + m_\alpha/c^2 + m_d/c^2 + m_t/c^2)/(m_\alpha/c^2) = 22.065$ MeV. For deuteron-triton c.m. energy $E$, corresponding deuteron laboratory energy $E_d = E \cdot (m_d/c^2 + m_t/c^2)/(m_t/c^2)$, neutron laboratory energy $E_n = E_{Th} + E_d \cdot (m_t/c^2)/(m_\alpha/c^2)$, and neutron-alpha c.m. energy $E_{n\alpha} = E_n \cdot (m_\alpha/c^2)/(m_\alpha/c^2 + m_n/c^2)$.

Expressions in Eq. (2) for Coulomb correction $C_0^2(\eta_d)$ and level shift $\Delta_d(E)$ that are dependent on Sommerfeld parameter $\eta_d$, are reproduced here, as reported in [23],

$$C_0^2(\eta_d) = 2\pi\eta_d(e^{2\pi\eta_d}-1)^{-1} \qquad \text{and} \qquad \Delta_d(E) = \frac{g_d^2}{\pi}\frac{\mu_d}{(\hbar c)^2 a_0}\left[-\gamma + \sum_{k=1}^{2^{10}}\left(\frac{1}{k}\cdot\frac{\eta_d^2}{k^2+\eta_d^2}\right) - \ln(\eta_d)\right] \qquad (4)$$

where the summation in $\Delta_d(E)$ over index $k$ is taken over $2^{10}$ terms, and where $\gamma = 0.5772157$ is the Euler-Mascheroni constant. Although not stated in [22] or [23], we find $0.003 \leq \Delta_d(E) \leq 0.657$ MeV over d-$^3$H c.m. energy range 0 to 0.3 MeV.

In Eq. (2), relativistic velocity $v_d$ in d-$^3$H c.m. system and Sommerfeld parameter $\eta_d$ are: $v_d = \frac{\hbar c^2 k_d}{\mu_d}$ and $\eta_d = \frac{1}{a_0 k_d}$ where the Bohr radius $a_0 = \frac{\hbar c}{\alpha \mu_d}$ and the fine structure constant $\alpha = 0.0072874$.

Relativistic velocity correction is less than 3x10$^{-6}$ at 0.3 MeV c.m. Difference of Sommerfeld parameter calculated using relativistic velocity via fine structure constant and via Bohr radius is constant and < 9x10$^{-6}$ over the entire energy range 0-0.3 MeV c.m.

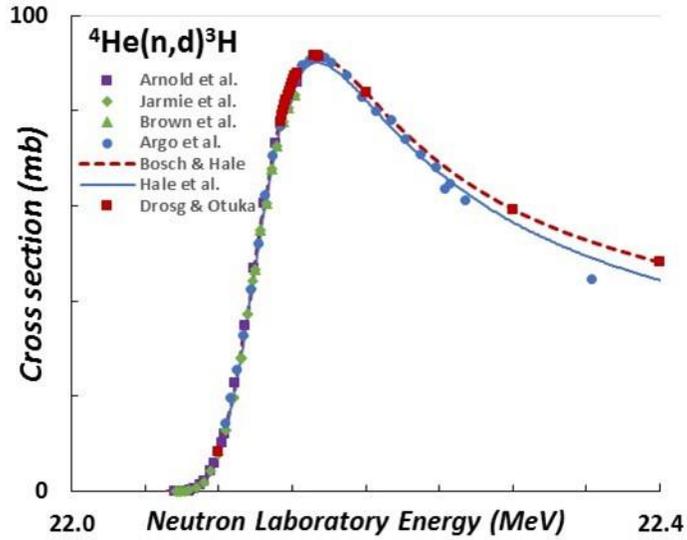

FIGURE 5: Cross section for $^4$He(n,d)$^3$H, determined by detailed balancing of cross section for $^3$H(d,n)$^4$He calculated from Eq. (2), is shown by solid (blue) curve. Dashed (red) curve, based on tabulated astrophysical S-factor for $^3$H(d,n)$^4$He reaction, represents $^3$H(d,n)$^4$He cross section reported by Bosch and Hale [24]. Red squares are from tabulation of Drosg and Otuka [25]. The (magenta) squares are from detailed balancing of data of Arnold et al. [26]; the (olive) diamonds are from data of Jarmie et al. [27]; the (green) triangles are from data of Brown et al. [28]; the (blue) circles are from data of Argo et al. [29].

In Fig. 5, at 22.19 MeV neutron laboratory energy (corresponding to 0.1 MeV d-$^3$H c.m.), $^4$He(n,d)$^3$H reaction cross calculated via detailed balancing from R-matrix fit of Bosch and Hale [24] (tabulated S-factor for $^3$H(d,n)$^4$He) is ca. 3% higher than effective field theory result. At 22.41 MeV neutron laboratory energy (corresponding to 0.275 MeV d-$^3$H c.m.), this difference grows to 9%. Fig. 4 of Hale, Brown, and Paris [23] indicate that the results of Bosch and Hale are close to the single-level fit at low energies, but the ratio to effective field theory result increases by about 5% at 0.1 MeV (d-$^3$H c.m.)

## Neutron-$^4$He elastic scattering cross section

The n-$^4$He elastic scattering cross section $\sigma_{elas}(E)$ vs neutron energy $E$ is taken as ([30], p 322)

$$\sigma_{elas}(E) = \sigma_{potential} + \sigma_{resonance} + \sigma_{interference} \tag{5}$$

where d-wave hard sphere (potential) scattering cross section is

$$\sigma_{potential} = \frac{4\pi}{k_n^2} g_c (2l+1) \sin^2 \phi, \tag{6}$$

elastic resonant cross section is

$$\sigma_{resonance} = \frac{\pi}{k_n^2} g_c \frac{\Gamma_n^2}{\left[(E-E_r)^2 + (\Gamma/2)^2\right]}, \tag{7}$$

and interference between potential and resonant scattering cross section is

$$\sigma_{interference} = \frac{\pi}{k_n^2} g_c \left[ \frac{2\Gamma_n(E-E_r)\sin 2\phi + \Gamma\Gamma_n(1-\cos 2\phi)}{(E-E_r)^2 + (\Gamma/2)^2} \right] \tag{8}$$

where $k_n^2$ is given in Eq. (1), statistical factor $g_c = 1.5$, and total width $\Gamma = \Gamma_n + \Gamma_d$.

## Background phase shift

Barker [12] reviews the sensitive dependence, in the one-level approximation, of the n-alpha total cross section on the value of the $d_{3/2}$ neutron background phase shift, which he calls $\phi$ as in [11], and which is just $-\phi_n$, the potential (hard sphere) phase shift.

For $l = 2$, the expression given in Table I, p. 1232 of Willard et al. [31], where for any argument $\alpha$, we substitute the identity, $\text{arccot } \alpha = \pi/2 - \arctan \alpha$

$$\phi_n(E) = k_n a_n - \frac{\pi}{2} - \arctan\left(\frac{(k_n a_n)^2 - 3}{3 k_n a_n}\right) \tag{9}$$

The expression Eq. (9) for $\phi_n$ confirms Barker's observation that for $a_n = 3$ fm [7, 27], $\phi_n \sim 29°$, for $a_n = 5$ fm [11], $\phi_n \sim 98°$ (not 5°), and that the equivalent value $\phi = 8° - 180° = -172°$ is the value of $-\phi_n$ for $a_n \sim 7$ fm. Barker also reviews justification for a "small" n-alpha channel radius of $a_n = 2.9$ fm originating in the work of Adair [32] and Dodder and Gammel [33].

## A model of n-$^4$He channel interaction radius

Let the n-$^4$He channel interaction radius $a_n(E)$ vary with energy $E$ (n-$^4$He c.m.) in the form of a Lorentz distribution,

$$a_n(E) = \frac{(1+\varepsilon) \cdot y_1}{y_0^2} - \frac{y_1}{(E-E_q)^2 + y_0^2}, \tag{10}$$

where $y_0$ is the width of the distribution centered on $E_q$, which is the energy at which $a_n(E_q)$ takes a minimum value $\varepsilon \cdot y_1 / y_0^2$, and where $y_1 / y_0^2$ is the value of $a_n(E)$ far from $E_q$. For $\varepsilon = 0.00001$, $y_0 = 0.1$ and $y_1 = 0.057$, the interaction radius $a_n = 5.7 \times 10^{-5}$ fm and 5.7 fm at $E_q$ and far from $E_q$, respectively.

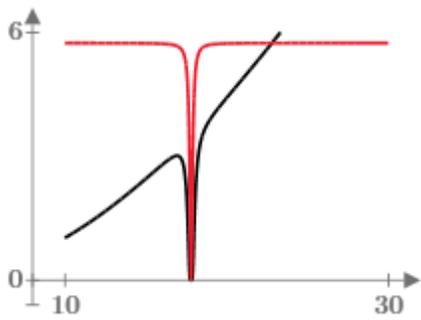

For $E_q$ = 17.802 MeV, interaction radius $a_n(E)$ in fm [Eq. (10), red curve] and corresponding background phase shift $\phi_n(E)$ in degrees [Eq. (9), black curve] using $a_n(E)$ from Eq. (10), vary with energy $E$ from 10 to 30 MeV, as shown in the adjacent plot.

It should be emphasized that the choice of $E_q$ =17.802 MeV as the energy at which $a_n$ reaches its minimum, corresponds to 0.179 ± 0.005 MeV in the d-$^3$H c.m. system, which is the energy eigenvalue of the $^5$He$^*$ unstable particle reported by Hale et al. [23]. These authors further remark that *"...no meaningful reduction in the $\chi^2$ was achieved by allowing separate values of the channel radii, so the fits were made with $a_d = a_n = a$...The best fit...was obtained for a = 7 fm, although $\chi^2$ was a shallow function of a in the range a = 3 to 8 fm."* As mentioned above, the value $a_n$ =5.7 fm far from $E_q$ is chosen in the present analysis.

**Time delay and speed**

Under assumption that interaction radius goes to zero at energy corresponding to formation of $^5$He$^*$ unstable particle, finite time delay and speed occur, as shown in Fig. 6. Scattering amplitude and phase shift for deuteron and neutron widths $\Gamma_d$ = 0.023 and $\Gamma_n$ = 0.064 MeV are shown on the left in Fig. 6.

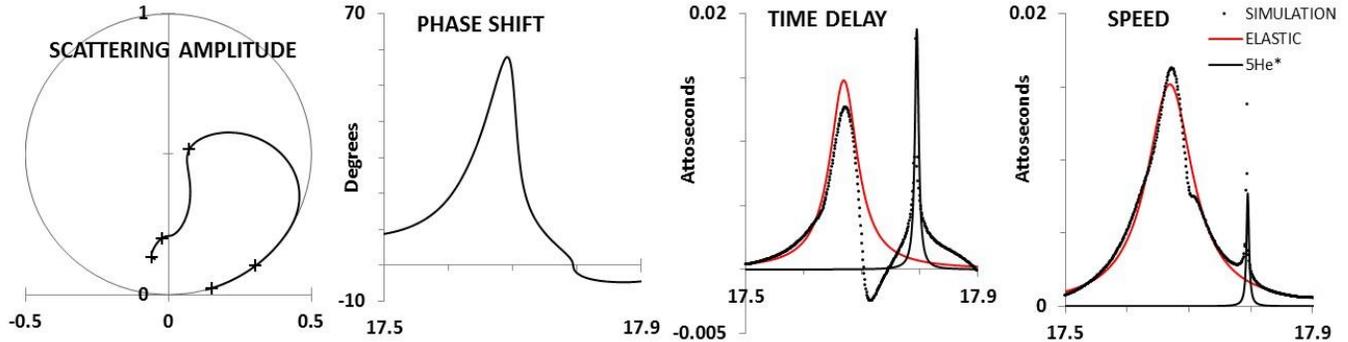

FIGURE 6: (Left two panels) D$_{3/2}$ scattering amplitude and corresponding phase shift. Scattering amplitude trajectory, calculated at 0.001-MeV steps from 17.50 to 17.90 MeV n-α c.m. energy does not encircle center of unitary circle. Large + symbols indicate values of n-α c.m. energy increasing (counterclockwise) in 0.1-MeV steps. Phase shift (in degrees) vs n-α c.m. energy (in MeV) does not increase through 90º, decreases and becomes negative at 17.795 MeV. (Right two panels) Corresponding real part of time delay and speed are compared with best fit to sum of two Lorentzians: elastic (red curve) and $^5$He$^*$ unstable particle (black curve). Peak time delay and speed occur at elastic resonance energy 17.669 MeV n-α c.m (22.129 MeV neutron laboratory energy) and at $^5$He$^*$ unstable particle energy 17.795 MeV (22.280 MeV neutron laboratory energy).

Using expressions for time delay and speed given elsewhere [15], we find values of time delay and speed of ca. 0.01 to 0.02 attosec for elastic resonance and for $^5$He$^*$ unstable particle, as shown on the right in Fig. 6. (N.B.: Time advance contribution associated with $^4$He(n,d)$^3$H reaction is evident but not shown in the time delay plot of Fig. 6.)

In summary, time delay and speed plot analyses, under the present model of n-α channel interaction radius that decreases to a value near zero at 17.802 MeV n-α c.m energy, provide further evidence of $^5$He$^*$ unstable particle formation at the energy of 17.795 MeV.

**E. Wigner leaps the mass 5 gap**

From: G. Gamow
*The Creation of the Universe*
The Viking Press, New York (1952)

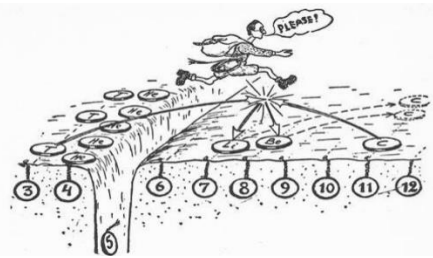